\documentstyle[epsfig,12pt]{article}

\def\be{\begin{equation}}
\def\ee{\end{equation}}

\def\d{\partial}
\def\eqref#1{(\ref{#1})}
\def\a{\alpha}
\def\bra{\langle}
\def\ket{\rangle}

\def\tr{{\rm tr}}
\def\kp{{k^\prime}}

\begin{document}

%----------------------------------------------------------------------
%                     T I T L E
%----------------------------------------------------------------------

\begin{titlepage}

\begin{centering}

\vspace*{3cm}

{\Large\bf Direct evidence for the Maldacena conjecture for 
${\cal N}=(8,8)$ super Yang--Mills theory in 1+1 dimensions}

\vspace{1.5cm}

{\bf John R. Hiller,$^a$ Stephen Pinsky,$^b$
Nathan Salwen,$^b$ and Uwe Trittmann$^c$}

\vspace{0.5cm}

{\sl ${}^a$Department of Physics \\
University of Minnesota Duluth\\
Duluth MN 55812}

\vspace{0.5cm}

{\sl ${}^b$Department of Physics \\
Ohio State University\\
Columbus OH 43210}

\vspace{0.5cm}

{\sl ${}^c$Department of Physics \& Astronomy\\
Otterbein College\\
Westerville OH 43081}

%\vspace{1.5cm}
%
%Version: June 27, 2005 

%----------------------------------------------------------------------%
%                       A B S T R A C T
%----------------------------------------------------------------------%

\vspace{1.5cm}

\begin{abstract}
We solve ${\cal N}=(8,8)$ super Yang--Mills theory in 1+1 dimensions
at strong coupling to directly confirm the predictions of supergravity 
at weak coupling. We do our calculations in the large-$N_c$ approximation 
using Supersymmetric Discrete Light-Cone Quantization with up to 
$3\times10^{12}$ basis states.  We calculate the stress-energy correlator
$\bra T^{++}(r) T^{++}(0) \ket$ as a function of the separation $r$ and 
find that at intermediate values of
$r$ the correlator behaves as $r^{-5}$ to within errors as predicted by 
weak-coupling supergravity. We also present an extension to significantly 
higher resolution of our earlier results for the
same correlator in the ${\cal N}=(2,2)$ theory and see that in this theory 
the correlator has very different behavior at intermediate values of $r$. 
\end{abstract}

\end{centering}

%\noindent
%PACS number(s):

\vfill

\end{titlepage}

\newpage

%----------------------------------------------------------------------%
%                       Introduction
%----------------------------------------------------------------------%

%%%%%%%%%%%%%%%%%%%%%%%%%%%%%%%%%%%%%%%%%%%%%%%%%%%%%%%%%%%%%%%%%%

\section{Introduction}

%%%%%%%%%%%%%%%%%%%%%%%%%%%%%%%%%%%%%%%%%%%%%%%%%%%%%%%%%%%%%%%%%%
We give a direct evidence of the duality between supersymmetric field
theory at strong coupling and supergravity at weak coupling. This 
conjectured duality, known as the Maldacena 
conjecture~\cite{Maldacena:1997re}, has been studied extensively, and 
there is considerable qualitative support for it. Most, if not all, of
these tests are in some way indirect, because of the difficulty of solving 
supersymmetric field theories at strong coupling.  The result we present 
here is unique, in that it is based on a numerical solution of an exactly 
supersymmetric field theory at strong
coupling.  

Our method is Supersymmetric Discrete Light-Cone
Quantization (SDLCQ)~\cite{sakai95,Lunin:1999ib}.
It is a well established tool for calculations of
physical quantities in supersymmetric gauge theory and has been exploited
for many supersymmetric Yang--Mills (SYM) theories. Briefly, the SDLCQ 
method rests on the ability to produce an exact representation of the 
superalgebra but is otherwise very similar to Discrete Light-Cone 
Quantization (DLCQ)~\cite{pb85,bpp98}. In DLCQ, we compactify the 
$x^-\equiv(t-z)/\sqrt{2}$ direction by putting the system on a circle with a 
period of $2L$, which discretizes the longitudinal momentum as $p^+=n\pi/L$,
with $n$ a positive integer. The total longitudinal momentum $P^+$ becomes
$K\pi/L$, where $K$ is an integer known as the harmonic 
resolution~\cite{pb85}. The positivity of the light-cone longitudinal 
momenta then limits the number of possible Fock states for a given $K$, 
and, thus, the dimension of Fock space becomes finite, enabling us to do 
some numerical computations. It is assumed that as $K$ approaches infinity, 
the solutions to this large, finite problem approach the solutions of the 
field theory. The difference between DLCQ and
SDLCQ lies in the choice of discretizing either $P^-$ or $Q^-$ to
construct the matrix approximation to the eigenvalue problem
$M^2|\Psi\ket=2P^+P^- |\Psi\ket=2P^+(Q^-)^2/\sqrt 2|\Psi\ket$, with
$P^+=K\pi/L$. For more details and additional discussion of SDLCQ, we
refer the reader to Ref.~\cite{Lunin:1999ib}.

The ${\cal N}=(8,8)$ supersymmetric theory in 1+1 dimensions in the 
large-$N_c$ limit is discussed in 
Refs.~\cite{Antonuccio:1998tm,Antonuccio:1999iz,Hiller:2000nf}; 
however, the published results are primitive compared to what can be 
obtained today, because of our greatly improved hardware and software.  
In the present paper we are able to present results for a resolution 
of $K=11$. At this resolution the problem has $3\times10^{12}$ basis 
states. By fully exploiting the symmetries of this theory,
we only need to deal with a subset of $3\times10^{7}$ such states.

We will look at the two-point correlation function of the
stress-energy tensor, namely  $\bra T^{++}(r)T^{++}(0)\ket$. The expected
behavior in the ultraviolet (UV) and infrared (IR) regions is $1/r^4$, 
and the predicted behavior in weak-coupling supergravity theory 
in the intermediate region is~\cite{Hashimoto:1999xu} $1/r^5$. We find 
the power behavior in ${\cal N}=(8,8)$ theory  
in the intermediate-$r$ region to be consistent with $1/r^5$.

An interesting effect in the calculation is that the 
finite-dimensional representations in SDLCQ with
odd and even values of $K$ result in very distinct solutions
of the SYM theory. We saw this result in our work on ${\cal N}=(2,2)$ 
SYM theory in two dimensions~\cite{Harada:2004ck}. Only at infinite 
resolution are the solutions identical.  
One might initially think that
having  two numerical representations of the supersymmetry is a
shortcoming of the SDLCQ approach, but it turns out to be an advantage
because it provides an internal measure of convergence.
We present new
results for the ${\cal N}=(2,2)$ theory, up to resolution $K=14$, 
which clearly demonstrate this. 

These results for ${\cal N}=(2,2)$ 
also correct a sign error in our earlier numerical results for 
${\cal N}=(2,2)$. 
The quantitative change is very small and has no qualitative effect.

The structure of this paper is the following. In Sec.~\ref{sec:N2SYM} we
review ${\cal N}$=(8,8) SYM theory. In Sec.~\ref{sec:sym} we discuss the 
symmetries of this theory, which play a critical role in allowing us to 
greatly reduce the basis set we need to retain. We discuss the two-point 
correlation function of the stress-energy tensor in Sec.~\ref{sec:cor}. 
Finally, in Sec.~\ref{sec:num}, we present our numerical results. A 
summary and some additional discussion are given in
Sec.~\ref{sec:discussion}.

%-----------------------------------------------------------------

\section{Review of ${\cal N}=(8,8)$ SYM Theory} \label{sec:N2SYM}

%-----------------------------------------------------------------

Before giving the numerical results, let us quickly review some
analytical work on ${\cal N}=(8,8)$ SYM theory, for the sake of 
completeness. For more details, see Ref.~\cite{Antonuccio:1998tm}.
This theory is obtained by dimensionally reducing ${\cal N}=1$ SYM
theory from ten dimensions to two dimensions. In light-cone gauge, where
$A_-=0$, we find for the action
\begin{eqnarray}
S^{LC}_{1+1}&=&\int dx^+ dx^- \tr \Bigg[ \d_+ X_I \d_-X_I
 +i\theta^T_R \d^+\theta_R+i\theta^T_L\d^-\theta_L  \\
 &&\quad +\frac 12 (\d_-A_+)^2+gA_+J^++\sqrt 2 g\theta^T_L\beta_I
 [X_I,\theta_R]+\frac {g^2}4 [X_I,X_J]^2 \Bigg], \nonumber
\end{eqnarray}
where $x^{\pm}$ are the light-cone coordinates in two dimensions,
the trace is taken over the color indices, the $X_I$ with $I=1,\ldots,8$ are
the scalar remnants of the transverse components of the
ten-dimensional gauge field $A_{\mu}$, the two-component spinor fields
$\theta_R$ and $\theta_L$ are remnants of the right-moving and left-moving
projections of the sixteen-component spinor in the ten-dimensional theory, 
and $g$ is the coupling constant. We also define the current
$J^+=i[X_I,\d_-X_I]+2\theta^T_R\theta_R$
and use the matrices $\beta_i$ given in Ref.~\cite{GreenSchwarzWitten}.

After using the equations of motion to eliminate all the non-dynamical 
fields, we find for $P^-=\int dx^- T^{+-}$ the expression
\be
P^-=g^2\int dx^- \tr\left(-\frac 12 J^+\frac 1{\d^2_-}J^+
   -\frac 14[X_I,X_J]^2+\frac i2 \beta_I[X_I,\theta_R])^T\frac 1{\d_-}
   \beta_J[X_J,\theta_R]\right).
\ee
The supercharges are found by dimensionally reducing the supercurrent in
the ten-dimensional theory. We find
\be
Q^-_{\a}=g\int dx^- \tr\left( -2^{3/4}J^+\frac
1{\d_-}u_{\a}+2^{-1/4}i[X_I,X_J](\beta_I\beta_J)_{\a\eta
}u_{\eta}\right),
\ee
where $\a,\eta=1,\ldots,8$ and the $u_{\a}$ are the components of $\theta_R$.
We expand the dynamical fields $X_I$ and $u_{\a}$ in Fourier modes as
\be X_{Ipq}(x^-)=\frac 1{\sqrt{2\pi}}\int_0^{\infty}
\frac {dk^+}{\sqrt{2k^+}}[A_{Ipq}(k^+)e^{-ik^+x^-}
 +A^{\dag}_{Iqp}(k^+)e^{ik^+x^-}],
\ee
\be
u_{\a pq}(x^-)=\frac 1{\sqrt{2\pi}}\int_0^{\infty}
   \frac {dk^+}{\sqrt{2}}[B_{\a pq}(k^+)e^{-ik^+x^-}
   +B^{\dag}_{\a qp}(k^+)e^{ik^+x^-}],
\ee
where $p,q=1,2,\ldots,N_c$ stand for the color indices, and
$A$ and $B$ satisfy the usual commutation relations
\begin{eqnarray}
[A_{Ipq}(k^+),A^{\dag}_{Jrs}(k^{'+})]
   &=&\delta_{IJ}\delta_{pr}\delta_{qs}\delta(k^+-k^{'+}), \\
\{B_{\a pq}(k^+),B^{\dag}_{\beta rs}(k^{'+})\}
   &=&\delta_{\a\beta}\delta_{pr}\delta_{qs}\delta(k^+-k^{'+}).
\end{eqnarray}

We work in a compactified $x^-$ direction of length $2L$ and ignore
zero modes.  With periodic boundary conditions, momenta are restricted
to a discrete set of values~\cite{sakai95}
$k^+ = {\pi} k/L$, with $k$ a positive integer.
The integrals over $k^+$ are replaced by sums: 
$\int dk^+\rightarrow\frac{\pi}{L}\sum_{k=1}^\infty$,
and Dirac delta functions become Kronecker deltas:
$\delta(k^+-k^{'+})\rightarrow\frac{L}{\pi}\delta_{kk'}$.
We then rescale the annihilation operators
\begin{equation}
\sqrt{\frac{L}{\pi}}a(k) = A(k^+ = \frac{\pi k}{L}), \;\;
\sqrt{\frac{L}{\pi}}b(k) =  B(k^+ = \frac{\pi k}{L}),
\end{equation}
so that
\begin{eqnarray}
  [a_{Ipq}(k),a^{\dag}_{Jrs}(\kp)]
   =\delta_{IJ}\delta_{pr}\delta_{qs}\delta_{k\kp}, \quad
   \{b_{\a pq}(k),b^{\dag}_{\beta rs}(\kp)\}
   =\delta_{\a\beta}\delta_{pr}\delta_{qs}\delta_{k\kp}.
\end{eqnarray}
The expansions become
\be
\label{eq:discretized}
X_{Ipq}(x^-)=\frac 1{\sqrt{2\pi}}\sum_{k=1}^{\infty}
\frac {1}{\sqrt{2k}}[a_{Ipq}(k)e^{-i\frac{\pi}{L}kx^-}
 +a^{\dag}_{Iqp}(k^+)e^{i\frac{\pi}{L}kx^-}],
\ee
\be
\label{eq:discretized2}
u_{\a pq}(x^-)=\frac 1{\sqrt{2L}}\sum_{k=1}^{\infty}
   \frac {1}{\sqrt{2}}[b_{\a pq}(k)e^{-i\frac{\pi}{L}kx}
   +b^{\dag}_{\a qp}(k)e^{i\frac{\pi}{L}kx^-}].
\ee
In terms of $a$ and $b$, the supercharge is given by
\begin{eqnarray}
   Q_{\a}^-  
&=&\frac{i2^{-1/4}g}{\pi}\sqrt{\frac{L}{\pi}}\sum^{\infty}_{k_1,k_2,k_3
   = 1}
    \delta_{(k_1+k_2),k_3} \Biggl\{  \\
   &&\times \Biggl[ \frac 1{2\sqrt{k_1k_2}}\left(\frac
{k_2-k_1}{k_3}\right)
    [b^{\dag}_{\a ij}(k_3)a_{Iim}(k_1)a_{Imj}(k_2)-a^{\dag}_{Iim}(k_1)
    a^{\dag}_{Imj}(k_2)b_{\a ij}(k_3)] \nonumber \\
   &&+ \frac 1{2\sqrt{k_1k_3}}\left(\frac {k_1+k_3}{k_2}\right)
    [a^{\dag}_{Iim}(k_1)b^{\dag}_{\a mj}(k_2)a_{I ij}(k_3)-a^{\dag}_{I
ij}
    (k_3)a_{Iim}(k_1)b_{\a mj}(k_2)] \nonumber \\
   &&+ \frac 1{2\sqrt{k_2k_3}}\left(\frac {k_2+k_3}{k_1}\right)
    [a^{\dag}_{I ij}(k_3)b_{\a im}(k_1)a_{Imj}(k_2)-b^{\dag}_{\a
im}(k_1)
    a^{\dag}_{Imj}(k_2)a_{I ij}(k_3)] \nonumber \\
   &&-\frac 1{k_1}[b^{\dag}_{\eta ij}(k_3)b_{\a im}(k_1)b_{\eta
mj}(k_2)
    +b^{\dag}_{\a im}(k_1)b^{\dag}_{\eta mj}(k_2)b_{\eta ij}(k_3)]
    \nonumber \\
   &&-\frac 1{k_2}[b^{\dag}_{\eta ij}(k_3)b_{\eta im}(k_1)b_{\a
mj}(k_2)
    +b^{\dag}_{\eta im}(k_1)b^{\dag}_{\a mj}(k_2)b_{\eta ij}(k_3)]
    \nonumber \\
   &&+\frac 1{k_3}[b^{\dag}_{\a ij}(k_3)b_{\eta im}(k_1)b_{\eta
mj}(k_2)
    +b^{\dag}_{\eta im}(k_1)b^{\dag}_{\eta mj}(k_2)b_{\a ij}(k_3)]
    \Biggr] \nonumber \\
    && + 2 
    \Biggl( \frac 1{4\sqrt{k_1k_2}}
    [b^{\dag}_{\alpha ij}(k_3)a_{I im}(k_1)a_{I mj}(k_2)
    +a^{\dag}_{I im}(k_1)a^{\dag}_{I mj}(k_2)b_{\alpha ij}(k_3)]
\nonumber
\\
   &&+\frac 1{4\sqrt{k_2k_3}}
    [a^{\dag}_{I ij}(k_3)b_{\alpha im}(k_1)a_{I mj}(k_2)
    +b^{\dag}_{\alpha im}(k_1)a^{\dag}_{I mj}(k_2)a_{I ij}(k_3)]
\nonumber
\\
   &&+\frac 1{4\sqrt{k_3k_1}}
    [a^{\dag}_{I ij}(k_3)a_{I im}(k_1)b_{\alpha mj}(k_2)
    +a^{\dag}_{I im}(k_1)b^{\dag}_{\alpha mj}(k_2)a_{I ij}(k_3)]
    \Biggr)\Biggr\},\nonumber
\end{eqnarray}   
where we have used the relation
$([\beta_I,\beta_J])_{\alpha,\eta} =\delta_{\alpha\eta} \delta_{IJ}$.

%%%%%%%%%%%%%%%%%%%%%%%%%%%%%%%%%%%%%%%%%%%%%%%%%%%%%%%%%%%%%%%%%%%%%%%%%%%

\section{Symmetries}

\label{sec:sym}

%%%%%%%%%%%%%%%%%%%%%%%%%%%%%%%%%%%%%%%%%%%%%%%%%%%%%%%%%%%%%%%%%%%%%%%%%%%

The superalgebra relations that involve $Q^+_{\a}$,
as specified by the anticommutators
\begin{equation}
\{Q^{+}_{\a},Q^{+}_{\beta}\}=\delta_{\a\beta}2\sqrt 2 P^{+}, \quad
   \{Q^+_{\a},Q^-_{\beta}\}=0,
\end{equation}
are satisfied, but the anticommutators for $Q_\alpha^-$
\begin{equation}  
\{Q^{-}_{\a},Q^{-}_{\beta}\}=\delta_{\a\beta}2\sqrt 2 P^{-}
\end{equation}
become in SDLCQ
\be
\{Q^{-}_{\a},Q^{-}_{\beta}\}\ne 0 \ {\rm if} \ \a\ne\beta, \qquad
\{Q^{-}_{\a},Q^{-}_{\a}\}=2\sqrt{2}P_\a^- .
\ee
Although we have different $P^-_{\a}$ for different
$Q^-_{\a}$, we can find unitary, self-adjoint transformations
$C_{\a'\a}$, such that $C_{\a'\a} P^-_{\a} C_{\a'\a} = P^-_{\a^\prime}$.  
Thus the eigenvalues of the different $P^-_{\a}$ are the same, and we
may choose any one of the $Q^-_{\a}$'s, at least for our
purposes.  In what follows we will use $Q^-_8$ and will
suppress the subscript unless it is needed for clarity.

To reduce the size of the numerical calculation, we seek symmetries that block
diagonalize the $P_8^-$ matrix.  One such symmetry is a $Z_2$ symmetry of
$Q_8^-$, called $S$-symmetry~\cite{Kutasov:1993gq}: 
$a_{I ij} \rightarrow -a_{I ji}$, $b_{\a ij}\rightarrow -b_{\a ji} $.
Extending the work of~\cite{Hiller:2000nf}, we also look for permutations
and sign changes of the bosonic and fermionic operators that leave
$Q_8^-$ unchanged. The terms not involving $\beta$ matrices are
preserved under all interchanges that fix $b_8$.  The additional
condition is that
$(\beta_I \beta_J^T - \beta_J \beta_I^T )_{\alpha\beta}
  b^{\dagger}_{\beta}a_{I }a_{J }$ be invariant with respect to
the replacement
\begin{eqnarray}
  a_{I} \rightarrow r(I)a_{p(I)},  \qquad
  b_{\beta} \rightarrow s(\beta)b_{q(\beta)}.
\end{eqnarray}
Defining $N_\alpha(\beta,I,J) =  (\beta_I \beta_J^T - \beta_J
\beta_I^T )_{\alpha\beta}$, this gives us the condition
\begin{eqnarray}
  N_\alpha(\beta,I,J)b^{\dagger}_{\beta}a_{I }a_{J }
  =  s(\beta) r(I)r(J)
N_\alpha(\beta,I,J)b^{\dagger}_{q(\beta)}a_{p(I)}a_{p(J)}.
\end{eqnarray}
Each possible flavor combination appears at most once on each side, 
so that we can set the coefficients equal, term by term, and obtain
the condition
\begin{eqnarray} \label{eq:FullCondition}
  N_\alpha(q(\beta),p(I),p(J))  =  s(\beta) r(I)r(J) N_\alpha(\beta,I,J).
\end{eqnarray}
This is then used to determine the allowed transformations
as specified by $r$, $s$, $p$, and $q$.
There are $(8!)^2 2^{16} $ such transformations.  When we
restrict these to transformations that leave $b_8$ unchanged, and thus have
$q(8) =8$ and $s(8) = 1$, we have instead a set of 
$8! 7! 2^{15} \approx 6.7 \times 10^{12} $ transformations.  

To determine the allowed transformations, we first 
find the permutations such that
\begin{eqnarray} \label{eq:AbsCondition}
  |N_\alpha(q(\beta),p(I),p(J))|  =  |N_\alpha(\beta,I,J)|.
\end{eqnarray}
Then, among these we check for choices of $r(I)$ and $s(I)$ that 
allow the full condition (\ref{eq:FullCondition}) to be satisfied.
There are 1344 permutations that satisfy the absolute-value 
condition (\ref{eq:AbsCondition}), and, for each of
these, there are 16 choices of $r(I)$ and $s(I)$ that satisfy the
full condition (\ref{eq:FullCondition}).
{F}rom these, we find that the group of transformations that leave $Q^-_8$ 
unchanged can be generated by 7 operators that square to $1$.  These 
generators are listed in Table~\ref{tab:generators}.

\begin{table}
\caption{Generators of transformations that leave $Q_8^-$ unchanged.
The first three transformations are permutations with sign changes, and 
the last four involve only sign changes.}
\label{tab:generators}
{\footnotesize
\begin{center}
\begin{tabular}{rrrrrrrrrrrrrrrr}
\hline \hline
   & $a_1$ & $a_2$ & $a_3$ & $a_4$ & $a_5$ & $a_6$ & $a_7$ & $a_8$ 
   & $b_1$ & $b_2$ & $b_3$ & $b_4$ & $b_5$ & $b_6$ & $b_7$ \\
\hline
1  & $a_1$ & $a_8$ & $-a_5$ & $-a_4$ & $-a_3$ & $a_6$ & $-a_7$ & $a_2$ 
   & $b_1$ & $b_4$ & $-b_3$ & $b_2$ & $b_7$ & $-b_6$ & $b_5$ \\
2  & $a_2$ & $a_1$ & $-a_5$ & $-a_6$ & $-a_3$ & $-a_4$ & $-a_8$ & $-a_7$ 
   & $b_4$ & $b_3$ & $b_2$ & $b_1$ & $b_5$ & $-b_6$ & $-b_7$ \\
3  & $a_2$ & $a_1$ & $-a_6$ & $a_8$ & $a_7$ & $-a_3$ & $a_5$ & $a_4$ 
   & $b_1$ & $-b_2$ & $b_6$ & $b_5$ & $b_4$ & $b_3$ & $-b_7$ \\
4  & $-a_1$ & $-a_2$ & $-a_3$ & $-a_4$ & $-a_5$ & $-a_6$ & $-a_7$ & $-a_8$ 
   & $b_1$ & $b_2$ & $b_3$ & $b_4$ & $b_5$ & $b_6$ & $b_7$ \\
5  & $a_1$ & $a_2$ & $a_3$ & $-a_4$ & $-a_5$ & $a_6$ & $-a_7$ & $-a_8$ 
   & $b_1$ & $b_2$ & $-b_3$ & $-b_4$ & $-b_5$ & $-b_6$ & $b_7$ \\
6  & $-a_1$ & $a_2$ & $a_3$ & $-a_4$ & $a_5$ & $-a_6$ & $-a_7$ & $a_8$ 
   & $b_1$ & $-b_2$ & $b_3$ & $-b_4$ & $-b_5$ & $b_6$ & $-b_7$ \\
7  & $a_1$ & $-a_2$ & $a_3$ & $a_4$ & $-a_5$ & $-a_6$ & $-a_7$ & $a_8$ 
   & $-b_1$ & $b_2$ & $b_3$ & $-b_4$ & $b_5$ & $-b_6$ & $-b_7$ \\
\hline \hline
\end{tabular}
\end{center}
}
\end{table}

%%%%%%%%%%%%%%%%%%%%%%%%%%%%%%%%%%%%%%%%%%%%%%%%%%%%%%%%%%%%%%%%%%%%%%%%%%%

\section{Correlation functions}

\label{sec:cor}

%%%%%%%%%%%%%%%%%%%%%%%%%%%%%%%%%%%%%%%%%%%%%%%%%%%%%%%%%%%%%%%%%%%%%%%%%%%

One of the physical quantities that we can calculate nonperturbatively is the
two-point function of the stress-energy tensor. Previous calculations of
this correlator in this and other theories can be found
in~\cite{Antonuccio:1999iz,Hiller:2000nf,Hiller:2001qb}.
Ref.~\cite{Hiller:2000nf} gives results
for the theory considered here but only for resolutions
$K$ up to 6. We can now reach $K=11$, with its 
$3\times10^{12}$ basis states, by careful use of the 
symmetries discussed in the previous section.
The largest matrices we have to consider are
$3\times10^7$ by $3\times10^7$.

We find that there are distinct behaviors in the correlation function
for even and odd $K$.  This is a familiar aspect of SYM theories with
extended supersymmetry, and we have argued~\cite{Harada:2004ck} that 
we have two different classes of representations
at finite $K$, which become identical as $K \to \infty$.

Let us first recall that there is a duality that relates the results for
the two-point function in ${\cal N}=(8,8)$ SYM theory to the results in
string theory~\cite{Hiller:2000nf}.  The correlation function on the
string-theory side, which can be calculated with use of the supergravity
approximation, was presented in~\cite{Antonuccio:1999iz}, and we will
only quote the result here. The computation is essentially a generalization
of that given in~\cite{Gubser:1998bc,Witten:1998qj}. The main conclusion
on the supergravity side was reported in~\cite{Hashimoto:1999xu}. Up to a
numerical coefficient of order one, which we have suppressed, it was
found that
\begin{equation}
   \bra T^{++}(r) T^{++}(0)\ket=\frac {N_c^{3/2}}{g_{YM}r^5}.
   \label{two}
\end{equation}
This result passes the following important consistency test. The SYM
theory in two dimensions with 16 supercharges has conformal fixed points 
in both the UV and the IR regions, with central charges of order $N_c^2$ 
and $N_c$, respectively. Therefore, we expect the two-point function of the
stress-energy tensor to scale like $N_c^2/r^4$ and $N_c/r^4$ in the deep
UV and IR regions, respectively.  According to the analysis
of~\cite{Itzhaki:1998dd}, we expect to deviate from
these conformal behaviors and cross over to a regime where the
supergravity calculation can be trusted. The crossover occurs at
$r=1/g_{YM}\sqrt{N_c}$ and $r=\sqrt{N_c}/g_{YM}$. At these points, 
the $N_c$ scaling of \eqref{two}
and the conformal result match in the sense of
the correspondence principle~\cite{Horowitz:1996nw}.
We should note here that this property for the correlation functions is
expected {\em only} for ${\cal N}=(8,8)$ SYM theory. However,
it would be natural to expect some similarity between ${\cal N}=(8,8)$
and ${\cal N}=(2,2)$ theories.

We wish to compute the expression
$F(x^-,x^+)=\bra T^{++}(x^-,x^+) T^{++}(0,0)\ket$, where we fix the
total momentum in the $x^-$ direction.  It is more natural to compute 
the Fourier transform and express the transform in a spectrally decomposed
form~\cite{Antonuccio:1999iz,Hiller:2000nf}
%
%\be
\begin{eqnarray}
\tilde F(P_-,x^+)&=&\frac 1{2L}\bra{T^{++}}(P_-,x^+){T^{++}}(-P_-,0)\ket \\
  &=&\sum_i \frac 1{2L}\bra 0|{ T^{++}}(P_-,0)|i\ket e^{-iP^i_+x^+}
  \bra i|{ T^{++}}(-P_-,0)|0\ket.  \nonumber
\end{eqnarray}
%\ee
%
The position-space form of the correlation function is recovered by
Fourier transforming with respect to $P_-=P^+ =K\pi/L$. We can continue
to Euclidean space by taking $r=\sqrt{2x^+x^-}$ to be real. The result 
for the correlator of the stress-energy tensor was presented
in~\cite{Antonuccio:1999iz}, and we only quote the result here:
\begin{equation} 
  F(x^-,x^+) =
   \sum_i \Big|\frac L{\pi}\bra 0|T^{++}(K)|i\ket\Big|^2\left(
   \frac {x^+}{x^-}\right)^2 \frac{M_i^4}{8\pi^2K^3}K_4(M_i
\sqrt{2x^+x^-}),
   \label{cor}
\end{equation}
where $M_i$ is a mass eigenvalue and $K_4(M_ir)$ is the modified Bessel
function of order 4. In~\cite{Antonuccio:1999iz} we found that the
stress-energy operator is given by
\be
T^{++}(x^-,x^+) =\tr \left[(\d_- X^I)^2+\frac 12(iu^{\a}\d_-u^{\a}-i
   (\d_-u^{\a})u^{\a})\right].
\ee
When written in terms of the discretized creation
operators,  $a^\dag$ and $b^\dag$, as in 
Eqs.~(\ref{eq:discretized}) and (\ref{eq:discretized2}), we find
\begin{eqnarray} \label{eq:Tpp}
   T^{++}(K)|0\ket &=&\frac {\pi}{2L}\sum_{k=1}^{K-1} 
    \left[-\sqrt{k(K-k)} a^{\dag}_{Iij}(K-k)a^{\dag }_{Iji}(k)\right. \\
   &&\rule{1in}{0mm}+\left.\left(\frac K2-k\right)
   b^{\dag }_{\a ij}   (K-k)b^{\dag }_{\a ji}(k)\right]|0\ket.\nonumber
\end{eqnarray}

The matrix element $(L/\pi)\bra 0|T^{++}(K)|i\ket$ is independent of $L$
and can be substituted directly to give an explicit expression for the
two-point function. We see immediately that the correlator behaves like
$1/r^4$ at small $r$, for in that limit, it asymptotes to
\be
\label{eq:smallr}
\left(\frac
{x^-}{x^+}\right)^2F(x^-,x^+)=\frac{N_c^2(2n_b+n_f)}{4\pi^2r^4}
   \left(1-\frac 1K \right).
\ee
On the other hand, the contribution to the correlator from strictly
massless states is given by
\begin{equation}
 \left(\frac {x^-}{x^+}\right)^2F(x^-,x^+)=\sum_i\Big|\frac L{\pi}
   \bra 0|T^{++}(K)|i\ket \Big|^2_{M_i=0}\frac 6{K^3\pi^2r^4}.
\label{larger}
\end{equation}
That is to say, we would expect the correlator to behave like $1/r^4$ at
both small and large $r$, assuming massless states have non-zero
matrix elements.

The operator used for calculating the correlator, $T^{++}(K)$
as given in Eq.~(\ref{eq:Tpp}), is preserved under all of the
$Z_2$ symmetries discussed in Sec.~\ref{sec:sym}.  
Therefore, we only need to consider states in the singlet sector. 
Including factors of 2 for supersymmetry and the $S$ symmetry, the
dimension of the $Q^-$ matrix is thus reduced by a factor of
$86016$.  Because the dimension of the basis increases by a factor of 
almost 17 for each unit increment in $K$, symmetry allows us to 
increase the resolution by more than four without increasing
the size of the matrix to be diagonalized.

%-----------------------------------------------------------------

\section{Numerical results}
\label{sec:num}
%-----------------------------------------------------------------

\begin{figure}[ht]
\begin{center}
\begin{tabular}{cc}
\psfig{figure=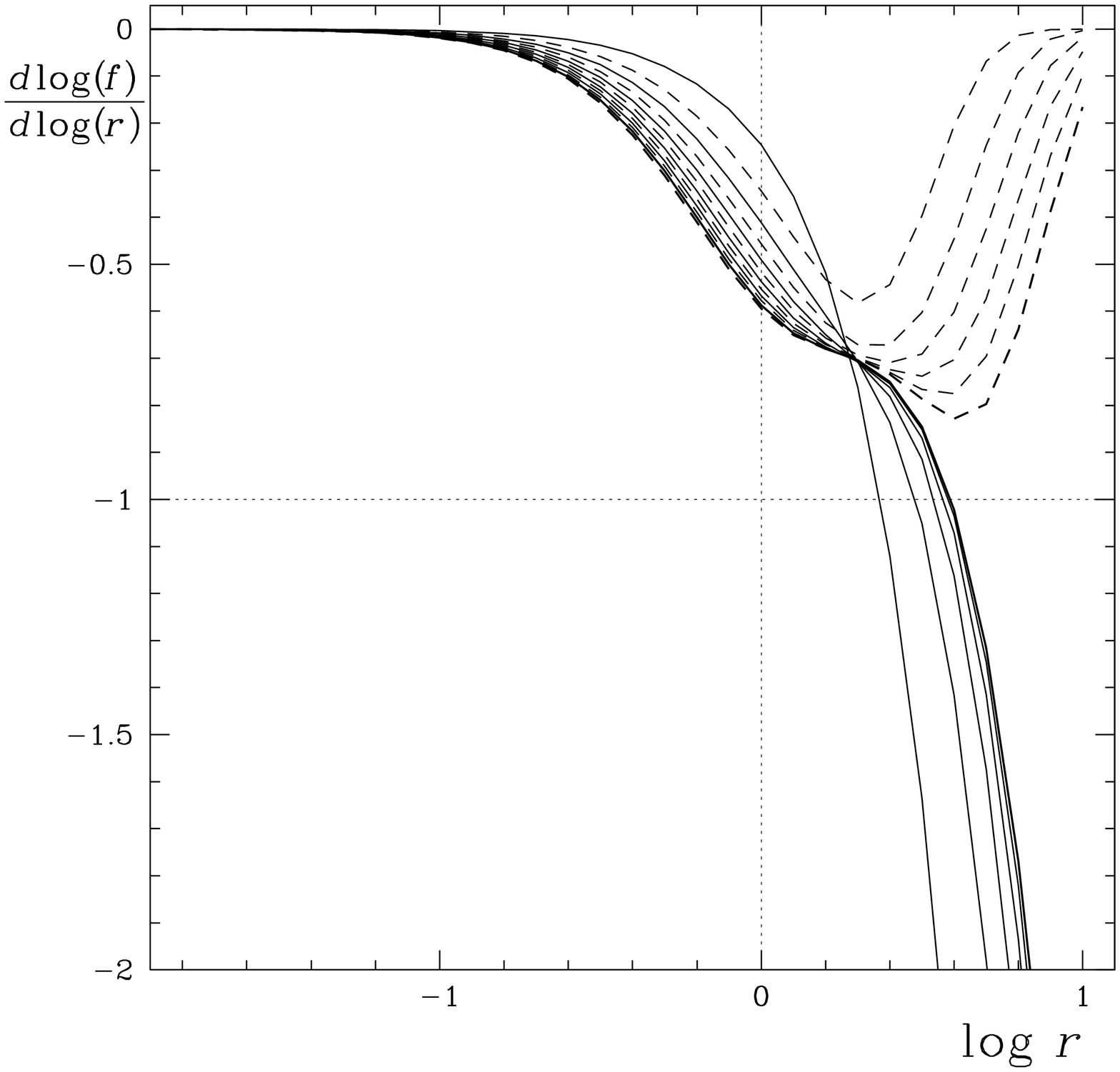,width=7 cm} 
& \psfig{figure=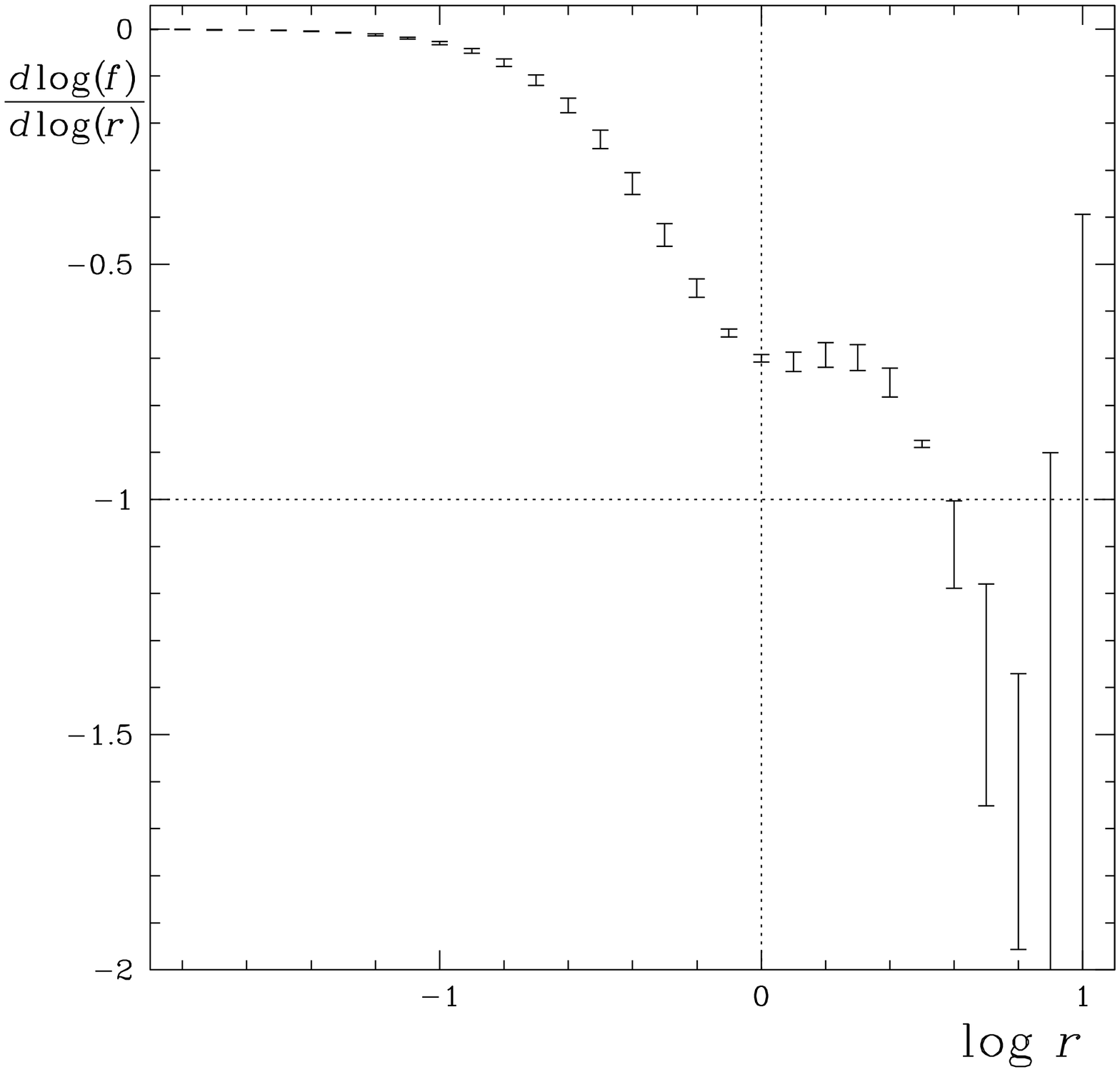,width=7 cm}   \\
(a)&(b) \\
\end{tabular}
\end{center}
\caption{The log-log derivative
$d\log_{10}(f)/d\log_{10}(r)$ of the scaled correlation function $f$, 
as defined in Eq.~(\ref{eq:Scaledf}) of the text, versus $\log_{10}(r)$,
for the ${\cal N} = (2,2)$ theory. The separation
$r$ is measured in units of $\sqrt{\pi/g^2N_c}$. 
In (a), the lines correspond to different values of resolution $K$
from 3 to 14, with dashed lines for even $K$ and solid for odd. The darker
lines are for $K=13$ and $K=14$; the lower-$K$ lines converge to these two.
In (b), the vertical bars span four extrapolations to infinite resolution
obtained from separate quadratic and cubic fits to even and odd $K$.}
\label{cor_Dcor22}
\end{figure}

\begin{figure}[ht]
\begin{center}
\begin{tabular}{cc}
\psfig{figure=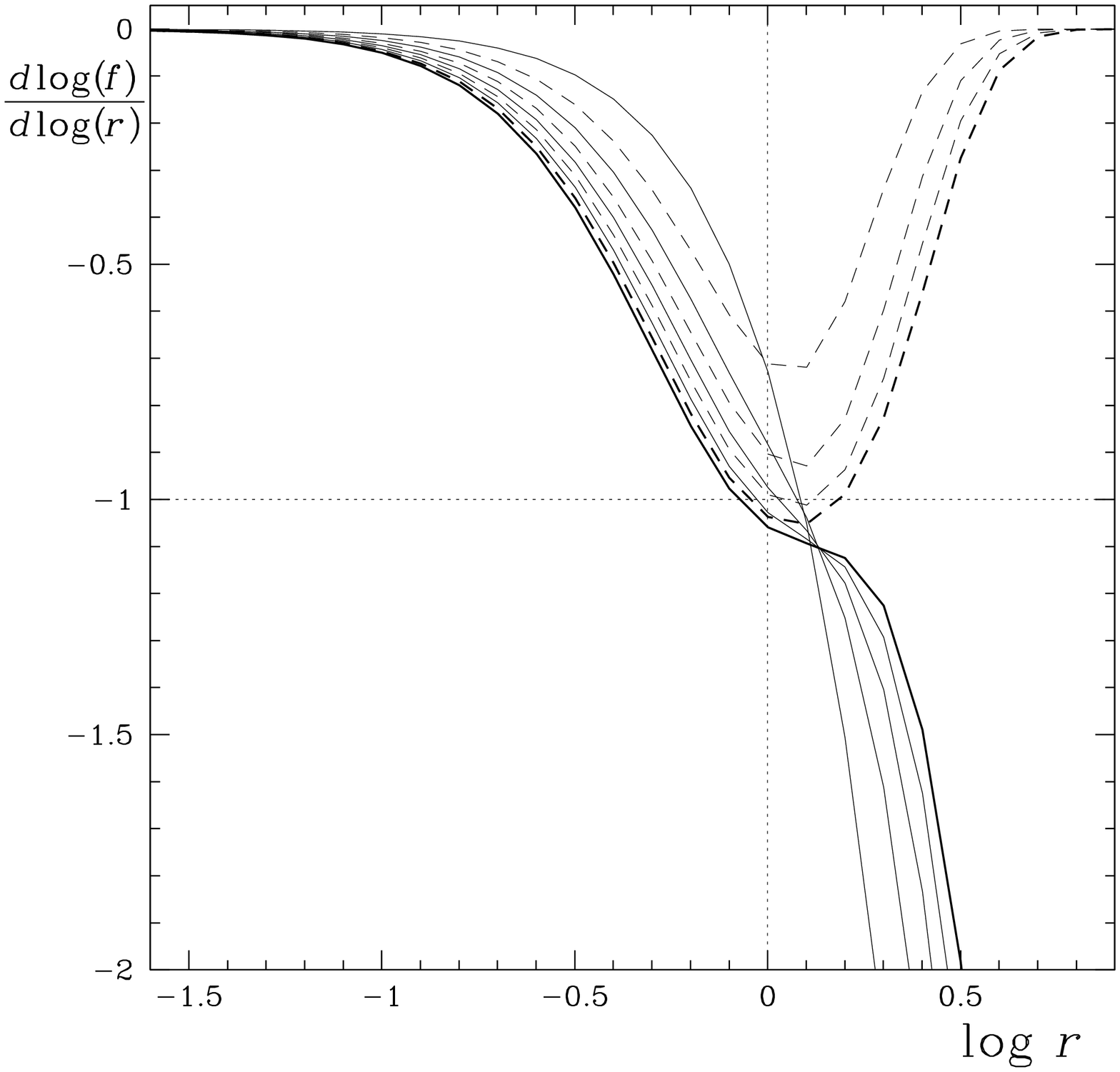,width=7 cm} 
& \psfig{figure=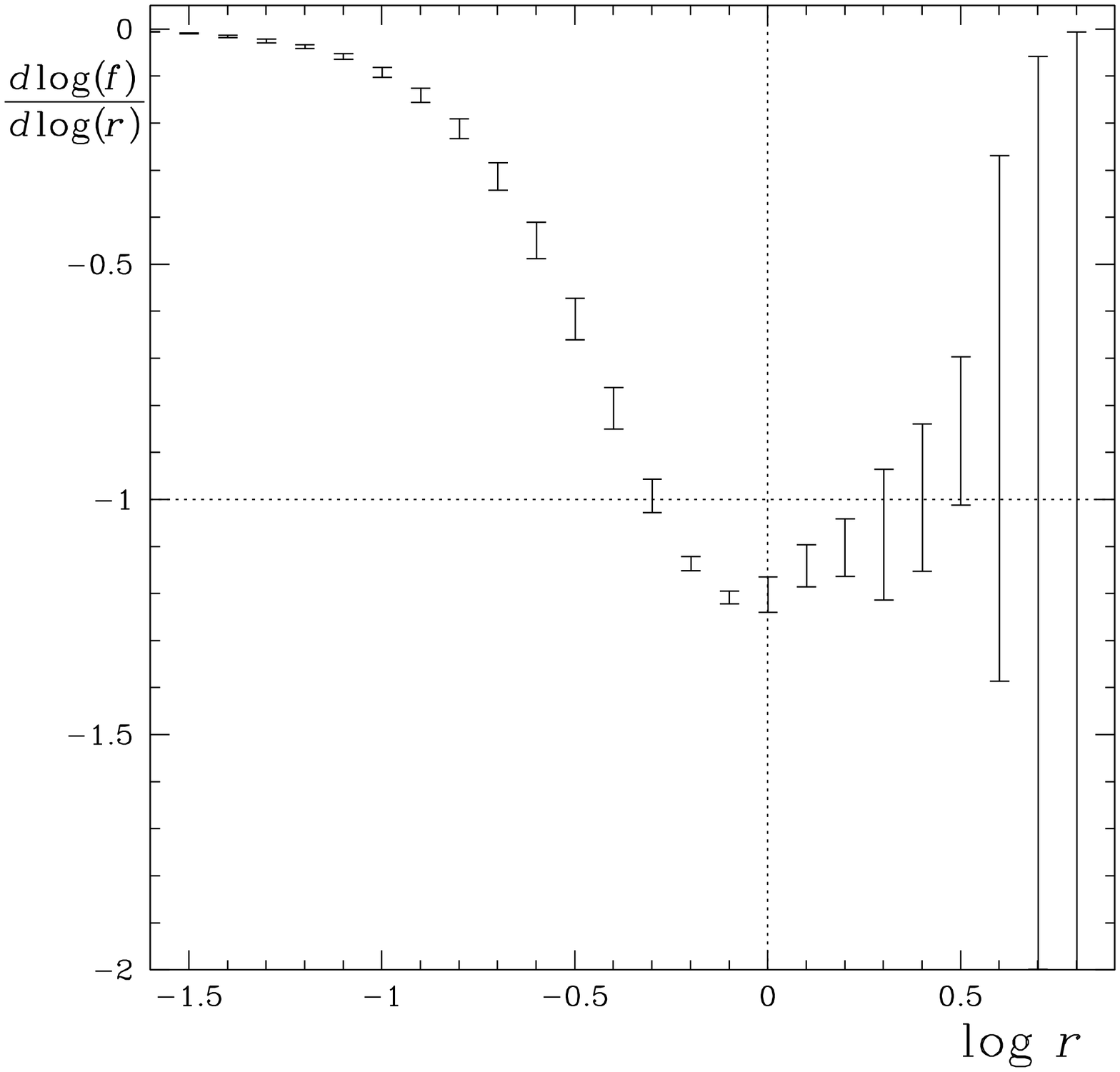,width=7 cm}   \\
(a)&(b) \\
\end{tabular}
\end{center}
\caption{Same as Fig.~\ref{cor_Dcor22}, except for the ${\cal N} = (8,8)$ 
theory with $K=10$ and 11 the largest values of the resolution.}
\label{cor_Dcor88}
\end{figure}

To compute the correlator using Eq.~\eqref{cor}, we approximate the sum
over eigenstates by a Lanczos~\cite{Lanczos}
iteration technique, as described in~\cite{Hiller:2000nf,Hiller:2001qb}.
The results are shown in 
Figs.~\ref{cor_Dcor22} and \ref{cor_Dcor88}, which show the log-log derivative 
$d\log_{10}(f)/d\log_{10}(r)$ of the scaled correlation function
\be \label{eq:Scaledf}
f\equiv \bra T^{++}(x^-,x^+)T^{++}(0)\ket
      \left(\frac{x^-}{x^+}\right)^2\frac{4\pi^2r^4}{N_c^2(2n_b+n_f)}.
\ee

\begin{figure}[ht]
\begin{center}
\begin{tabular}{cc}
\psfig{figure=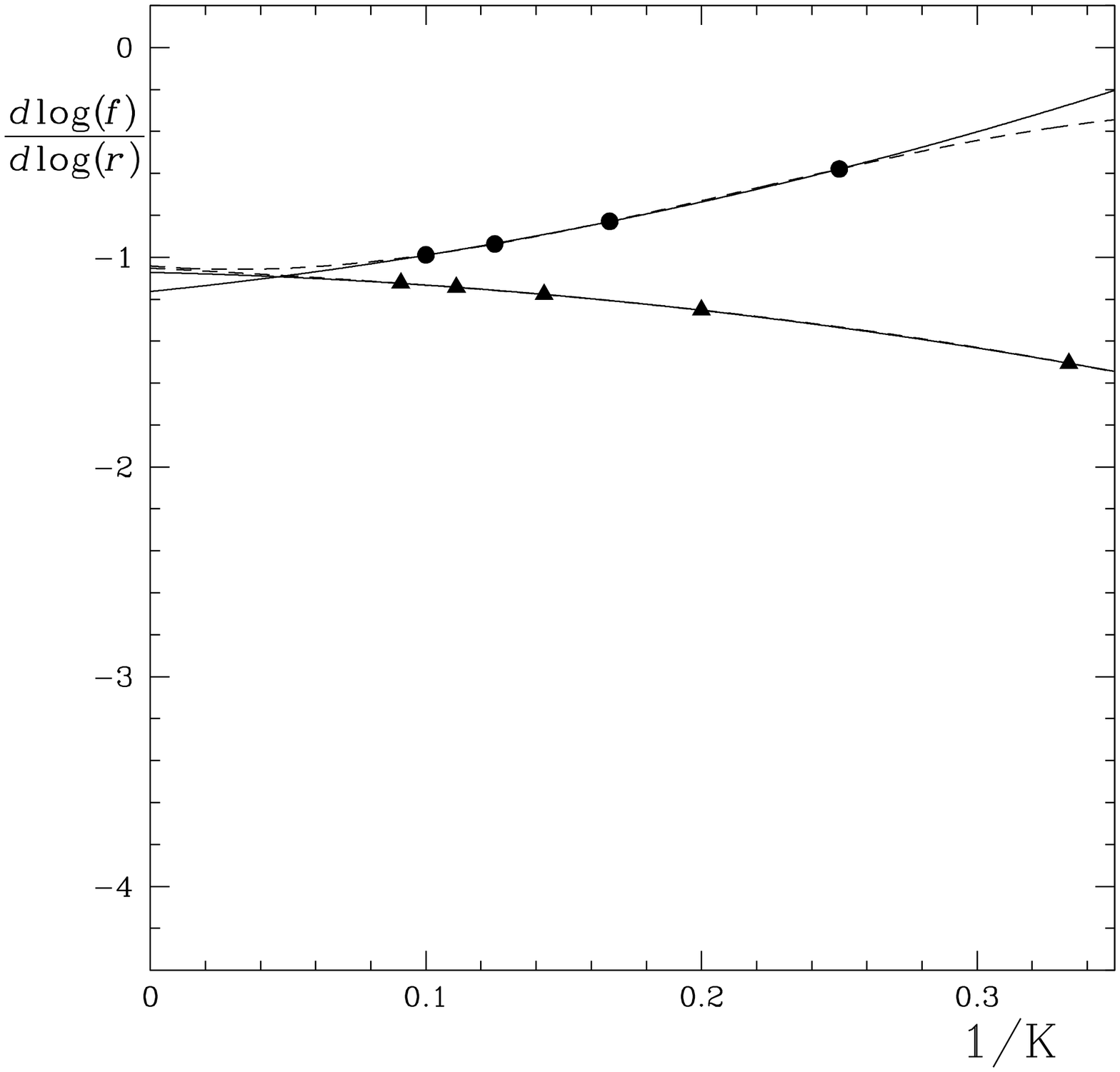,width=7 cm} 
& \psfig{figure=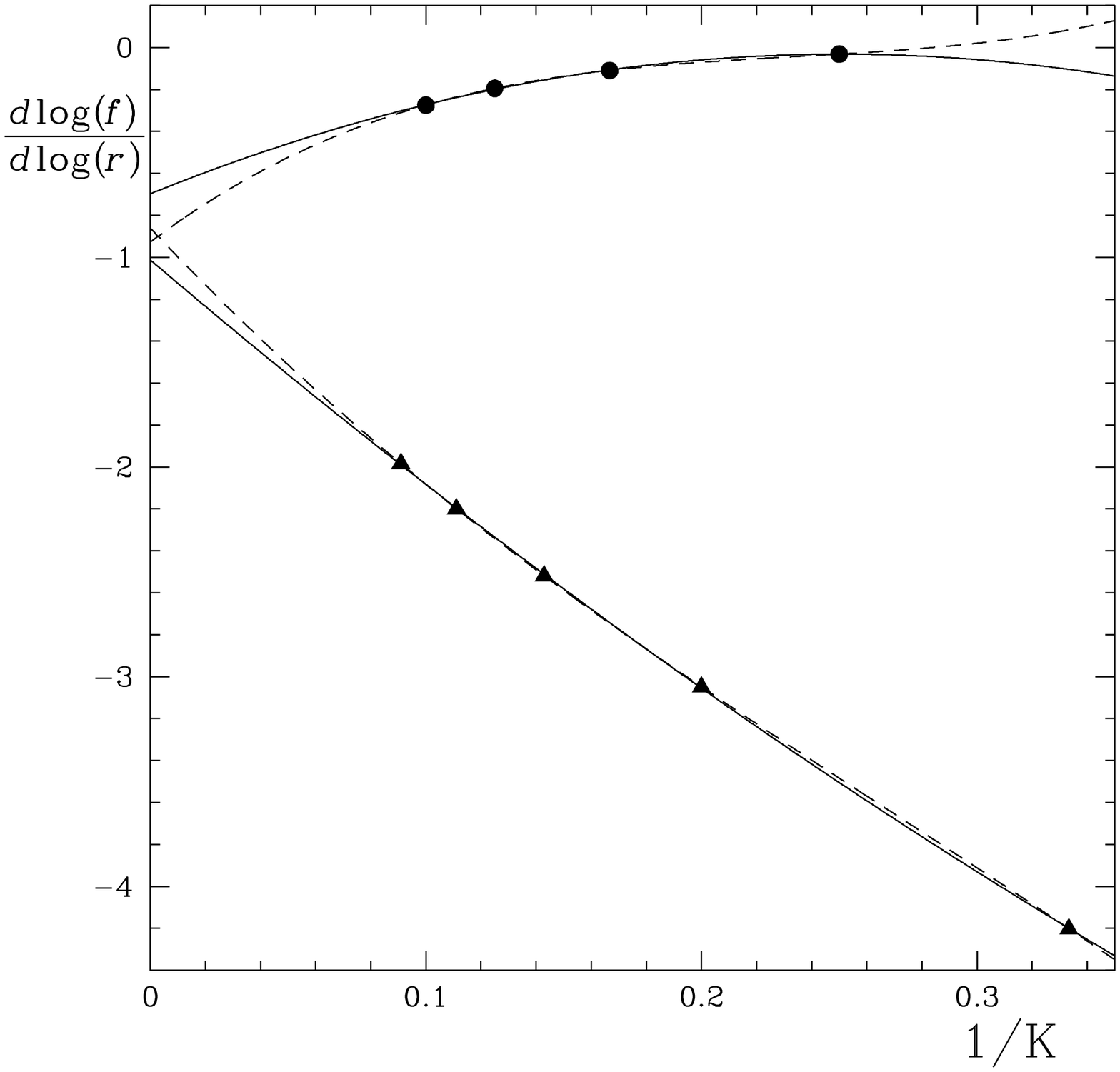,width=7 cm}   \\
(a)&(b) \\
\end{tabular}
\end{center}
\caption{Sample quadratic (solid) and cubic (dashed) fits for extrapolations 
to infinite resolution in the ${\cal N} = (8,8)$ theory.
The log-log derivative $d\log_{10}(f)/d\log_{10}(r)$ of the scaled 
correlation function $f$, as defined in Eq.~(\ref{eq:Scaledf}) of the text,  
is plotted versus $1/K$ for (a) $\log_{10}(r)=0.2$ and 
(b) $\log_{10}(r)=0.5$, where $r$ is measured in units of $\sqrt{\pi/g^2N_c}$.
Computed values of the log-log derivative are marked by circles for even $K$
and triangles for odd $K$.}
\label{cor_fit}
\end{figure}

Let us discuss the behavior of the correlator at small,
large, and intermediate $r$.
First, at small $r$, the graphs  of $f$ for different $K$ approach 0
as $K$ increases.  This follows Eq.~(\ref{eq:smallr}), which
gives the form $f = 1-1/K$.
Second, at large $r$, obviously, the behavior is different for even and
odd $K$ . However, the difference gets smaller as $K$ gets larger.
The reason for this difference is that there are different 
finite-dimensional representations of the super-algebra~\cite{Harada:2004ck} 
for even and odd $K$,  which only coincide at infinite
resolution. Looking at the detailed information of the computation of 
the correlator at larger $r$, we found that for even $K$ there are massless 
states that contribute to the
correlator, while there is no massless state that makes any contribution 
for odd $K$.  Instead, it is  the lowest massive states that contribute 
the most for odd $K$, and these
states become massless as the resolution approaches infinity.  

The results for the ${\cal N}=(2,2)$ theory are shown in Fig.~\ref{cor_Dcor22}, 
and  the results for the ${\cal N }=(8,8)$ theory in  
Fig.~\ref{cor_Dcor88}.  The results for
the ${\cal N}=(2,2)$ theory are obtained for higher resolution and 
more clearly show the convergence of the two representations. We are now 
able to extend our calculation  of the
${\cal N}=(2,2)$ SYM theory from $K=11$~\cite{Harada:2004ck} to $K=14$. To 
calculate continuum results in SDLCQ, it is customary to extrapolate  the 
finite resolution results to infinite
resolution.  We have done this for the curves for the ${\cal N}= (2,2)$ and  
$(8,8)$ theories by performing the extrapolation at successive values of $r$.
We fit the even and odd resolution points with separate quadratic and cubic 
functions of $1/K$ to perform four extrapolations.   Typical results of these 
calculations are shown in Fig.~\ref{cor_fit} for the ${\cal N}= (8,8)$ theory
at $\log_{10}(r)=0.2$ and at $\log_{10}(r)=0.5$, where $r$ is measured in 
units of $\sqrt{\pi/g^2N_c}$.

In the intermediate-$r$ region for the ${\cal N}=(2,2)$ theory, clearly the 
behavior of the correlation in Fig.~\ref{cor_Dcor22}(a) changes, and there 
is a flat region. The behavior is
even more apparent in Fig.~\ref{cor_Dcor22}(b), the extrapolation to infinite 
resolution.  In~\cite{Harada:2004ck} we first saw   the flat region for the
${\cal N}=(2,2)$ SYM theory, indicating that the correlator in this case 
behaves like $1/r^{4.7}$. Note that the region of flattening extends farther 
out as $K$ gets larger, for both odd and even $K$, implying again that the 
even and odd representations appear to agree as $K$ goes to infinity. 
Unfortunately, there is currently no known prediction for 
the behavior of the correlator  in the intermediate-$r$ region this theory 
or even any reason to believe that the correlator should have a special 
behavior at intermediate values of $r$. 

In the intermediate-$r$ region for the ${\cal N}$=(8,8) theory, there is a 
prediction that the correlation should have a special behavior. This predicted 
behavior is very different from the behavior we found in the 
${\cal N}=(2,2)$ SYM theory.  We  expect from
Eq.~\eqref{two} that the behavior is
$1/r^5$. In~\cite{Hiller:2000nf} we found that the correlator may be 
approaching  this behavior, and we indicated that 
conclusive evidence would be a flat region in the derivative of the  scaled 
correlator at a value of $-1$.  
In Fig.~\ref{cor_Dcor88}(b) we see that as $r$ increases, the slope 
approaches $-1$. It overshoots $-1$ and then appears to approach
$-1$ from below and to remain consistent with that value for an extended 
range of $r$. 
Beyond this point, we see that the errors increase significantly. 
Thus, there is a finite range of 
intermediate values of $r$ over which the numerical solution of 
strong-coupling SYM theory is consistent with the $1/r^5$ prediction of 
weak-coupling supergravity.

%%%%%%%%%%%%%%%%%%%%%%%%%%%%%%%%%%%%%%%%%%%%%%%%%%%%%%%%%%%%%%%%%%%%%%%%%%%%
%

\section{Discussion} \label{sec:discussion}

%%%%%%%%%%%%%%%%%%%%%%%%%%%%%%%%%%%%%%%%%%%%%%%%%%%%%%%%%%%%%%%%%%%%%%%%%%%%
%%

We have presented  numerical results for the two-point correlation
function of the stress-energy tensor, using SDLCQ in
$1+1$ dimensions in the large-$N_c$ approximation, for
${\cal N}=(2,2)$ SYM theory up to resolution $K=14$ and for  
${\cal N}=(8,8) $ SYM theory up to resolution $K=11$.  There are 
two distinct classes of representations for these SYM theories,
one for the resolution $K$ even and one for $K$ odd, and these
representations become identical as $K\to \infty$. The two-point 
correlators of the stress-energy tensor behave like
$1/r^4$ in the UV (small $r$) and IR (large $r$, $K$ even) regions.  
The large-$r$ behavior for $K$ odd, on the other hand, has an exponential
decay, since in the odd $K$ representations the massless states 
become massless only as $K\to\infty$.

While there is no known prediction for the correlator of the stress
energy tensor at intermediate values of $r$ for
the ${\cal N}=(2,2)$ theory, it is interesting that we find the 
dominant power law behavior $1/r^{4.7}$. There is clear convergence 
over a wide range of intermediate values of $r$. 

The results presented here also include  a correction to our earlier work on 
the ${\cal N}=(2,2)$ theory. We recently found a sign  error in one term in 
our numerical calculations.  This correction does not change the qualitative 
behavior found in our previous calculation; the quantitative change is
quite small.

In ${\cal N}=(8,8)$  SYM theory in $1+1$ dimensions, the 
correlator is expected to behave like $1/r^5$ in the intermediate region. 
We were able to confirm this power-law behavior with a flat region in the
derivative of the scaled correlator. We have done this by calculating the
correlator at successive values of the resolution and extrapolating the 
function to infinite resolution. 
The maximum resolution we reached was $K=11$. At 
that resolution the total number of basis
states in the SDLCQ approximation is $3\times10^{12}$. Using all of the
symmetries of the ${\cal N}=(8,8)$ theory, including some that are specific to 
the sector where the correlator is non-vanishing, we were able to reduce 
the number of basis states that we needed to consider to $3\times10^{7}$.

While just a few years ago it seemed inconceivable that one could reach
these resolutions in SDLCQ, advances in our understanding of the problem 
as well as advances in software and hardware have allowed us to do so. 
We remain confident that we can, in fact,
reach even higher resolutions. At this point we believe that the
most important direction is to increase the number of fields that we consider. 
In particular, we are interested in
considering theories with both extended supersymmetry and fundamental matter.

%%%%%%%%%%%%%%%%%%%%%%%%%%%%%%%%%%%%%%%%%%%%%%%%%%%%%%%%%%
\section*{Acknowledgments}
This work was supported in part by the U.S. Department of Energy
and the Minnesota Supercomputing Institute. One of the authors (U.T.)
would like to thank the Research Corporation for supporting his work.
%%%%%%%%%%%%%%%%%%%%%%%%%%%%%%%%%%%%%%%%%%%%%%%%%%%%%%%%%%

\end{document}